\documentclass[
  ,final            
  ]
  {aipproc}

\layoutstyle{6x9}

\begin{document}

\title{The Galaxy Proximity Effect in the Ly$\alpha$ Forest}

\author{Juna A. Kollmeier}{
  address={The Ohio State University, Dept. of Astronomy, Columbus, OH
43210} }
\author{David H. Weinberg}{
  address={The Ohio State University, Dept. of Astronomy, Columbus, OH 43210}
}

\author{Romeel Dav\'e}{
  address={University of Arizona, Dept. of Astronomy, Tucson, AZ 85721}
}
\author{Neal Katz}{
  address={University of Massachusetts, Dept. of Physics and Astronomy, Amherst, MA, 91003}
}
\def\ltorder{\mathrel{\raise.3ex\hbox{$<$}\mkern-14mu
             \lower0.6ex\hbox{$\sim$}}}
\def\gtorder{\mathrel{\raise.3ex\hbox{$>$}\mkern-14mu
             \lower0.6ex\hbox{$\sim$}}}

\begin{abstract}
 Hydrodynamic cosmological simulations predict that the average
 opacity of the Ly$\alpha$ forest should increase in the neighborhood
 of galaxies because galaxies form in dense environments.  Recent
 observations (Adelberger et al. \cite{adelberger02}) confirm this
 expectation at large scales, but they show a {\it decrease} of
 absorption at comoving separations $\Delta_r \ltorder 1 h^{-1}$ Mpc.  We
 show that this discrepancy is statistically significant, especially
 for the innermost data point at $\Delta_r \le 0.5 h^{-1}$ Mpc, even though
 this data point rests on three galaxy-quasar pairs.  Galaxy redshift
 errors of the expected magnitude are insufficient to resolve the
 conflict.  Peculiar velocities allow gas at comoving distances $\gtorder 1
 h^{-1}$ Mpc to produce saturated absorption at the galaxy redshift,
 putting stringent requirements on any ``feedback'' solution.  Local
 photoionization is insufficient, even if we allow for recurrent AGN
 activity that keeps the neutral hydrogen fraction below its
 equilibrium value.  A simple ``wind'' model that eliminates all
 neutral hydrogen in spheres around the observed galaxies can marginally explain
 the data, but only if the winds extend to comoving radii $\sim 1.5
 h^{-1}$ Mpc.
\end{abstract}

\maketitle


\section{Basic Predictions}

In a recent paper \cite{kollmeier02}, we discuss a variety of predictions for
galaxy-Ly$\alpha$ forest correlations from hydrodynamic simulations.
In this proceeding, we focus on the ``galaxy proximity'' effect on
small scales ($\le 2 h^{-1}$ Mpc comoving).  Using smoothed particle
hydrodynamics simulations (SPH) of a $\Lambda$CDM universe
($\Omega_m=0.4, \Omega_\Lambda=0.6, h=0.65,\Omega_b=0.02
h^{-2}=0.0473$, $\sigma_8=0.80$), we generate synthetic Ly$\alpha$
forest spectra from the temperature, gas density, and velocity at each
spatial location in skewers through the simulation box.  Once we have
created synthetic spectra, we use the known positions of the simulated
galaxies to compute the mean flux decrement, $\langle D\rangle =
\langle 1 - e^{-\tau}\rangle$, as a function of comoving separation
from the galaxy, $\Delta_r$.

Figure 1a shows the basic predictions for the mean flux decrement,
computed for the 150 galaxies with the highest star formation rates
within a simulation $50 h^{-1}$ Mpc on a side.  We see a clear trend
of increasing decrement (absorption) with decreasing $\Delta_r$, a
signature of the dense environments these galaxies occupy. The
observed points, from \cite{adelberger02}, show a similar trend at
large scales, but the trend flattens at $\Delta_r \ltorder 2.5 h^{-1}$ Mpc,
and absorption {\it decreases} for $\Delta_r \ltorder 1 h^{-1}$ Mpc.  The
innermost point, $0 \le \Delta_r \le 0.5 h^{-1}$ Mpc is in especially
severe conflict with the theoretical prediction.  We investigate
several possibilities for resolving this conflict below.  Similar
investigations have been carried out by \cite{croft02},
\cite{bruscoli02}, and our results are compatible with theirs to the
extent that they overlap.

\begin{figure}
  \includegraphics[height=.5\textheight]{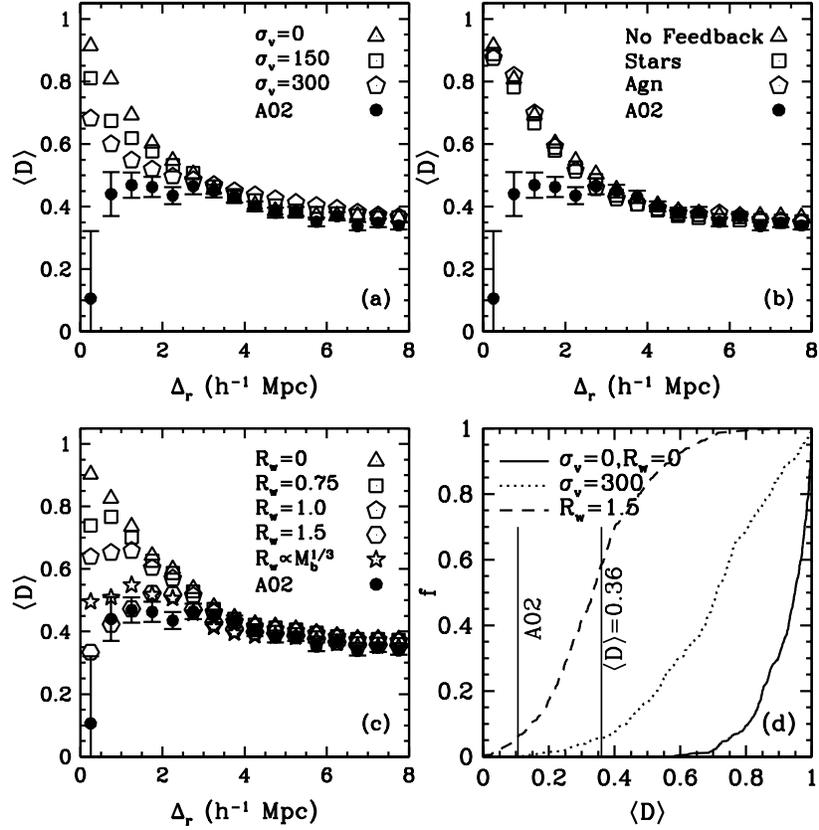} \caption{The
  Conditional Mean Flux Decrement.  (a) The conditional mean decrement
  compared with the baseline calculation, and calculations with
  redshift errors as indicated in panel and described in text, (b)
  Effect of photoionization feedback from stars or AGN on the
  numerical predictions, (c) Effect of spherical winds/gas removal on
  the predictions, (d) Cumulative distribution of 3-tuples at
  $\Delta_r=0.25$ Mpc as a function of $\langle D\rangle$. Vertical lines at $\langle
  D\rangle=0.11$ and $0.36$ correspond to the Adelberger et
  al. \cite{adelberger02} value for this separation and the global mean
  respectively. Symbols are as indicated in the panel.}
\end{figure}

\section{Redshift Errors}
Since we expect galaxies to occupy peaks in the density distribution,
if the redshift of the galaxy is incorrectly estimated, then one may
compute an artificially low value of $\langle D\rangle$ at small
$\Delta_r$ because one samples a region of lower density than the
region around the true galaxy redshift.  We estimate the size of this
effect by adding a redshift error, drawn from a Gaussian distribution,
to each galaxy redshift. Squares and pentagons in Figure 1a show the
result for rms redshift errors $\sigma_v=150 {\rm km}\;{\rm s}^{-1}$ and
$\sigma_v=300 \rm {km}\;{\rm s}^{-1}$, respectively.  The sign of the
effect is as expected --- increasing the redshift errors does decrease
the values of $\langle D\rangle$ at small $\Delta_r$, pushing them
towards the global mean.  It is clear, however, that even with
significant redshift errors (Adelberger et al. estimate $\sim150
{\rm km}\;{\rm s}^{-1}$) the difference between the theoretical curves
and the observations remains substantial.

\section{Local Photoionization}
There is some evidence that a significant amount of Lyman continuum
radiation is leaking from the interstellar media of Lyman Break
Galaxies (LBGs) \cite{steidel01}.  It is then plausible that these
galaxies may affect their immediate surroundings in the form of
photoionization from the stars within them.  There is also evidence
indicating that $\sim 3\%$ of LBGs host AGN \cite{steidel02}. If the
timescale between AGN outbursts is sufficiently short, then, in
contrast to the stellar case, the gas surrounding the galaxies can
remain out of photoionization equilibrium between outbursts, and the
neutral hydrogen fractions around galaxies that have recently hosted
AGN may be further suppressed.  We have incorporated simple models for
these two scenarios within the simulations by including the
non-uniform ionizing background near galaxies in each case, as well as
the additional effect of non-equilibrium neutral fractions in the AGN
case.  For details of these models see \cite{kollmeier02}.  Figure 1b
compares the results of these two calculations to the original,
no-feedback, calculation. Photoionization has minimal impact on the
mean decrement even at small $\Delta_r$. It is tempting to think that
increasing either the AGN luminosities or the escape fraction of
ionizing photons could produce a larger effect, but this is not the
case because the total output of the sources cannot exceed the UV
background, which is itself constrained by the mean (unconditional)
flux decrement.  The models we have presented are close to maximal,
with the observed galaxies or their AGN assumed to produce $50\%$ of
the entire UV background.

\section{Winds}
 Outflows have been detected in LBGs by looking at the difference
between absorption and emission features within these systems
\cite{pettini02}.  Strong winds from supernovae are a generic property
of starburst galaxies and have the effect of shocking and sweeping up
the material in their wake, both of which lead to decreased absorption
inside the ``sphere of influence'' of the wind.  We have constructed
very simple ``wind'' models in which we eliminate all neutral hydrogen
in a spherical region of radius $R_{wind}$ around each galaxy in the
simulation sample.  We note that this model is highly optimistic since
it assumes either {\it perfect} entrainment of the material in the
volume out to $R_{wind}$ or sufficient energy injection to completely
ionize hydrogen within this radius. Neither condition is necessarily
expected to hold for realistic winds (\cite{croft02},
\cite{bruscoli02}).

Figure 1c shows the result of the wind model for the 40 galaxies with
the highest star formation rates in a simulation box of side $22.22
h^{-1}$ Mpc (comoving).  Here the squares, pentagons, and hexagons
correspond to models with constant comoving radii $R_{wind}=0.75,
1.0$, and $1.5 h^{-1}$ Mpc respectively.  Stars show a model in which
the volume of the wind around a galaxy is proportional to the baryonic
mass of the galaxy, normalized such that the $40th$ brightest galaxy
has a wind radius of $1 h^{-1}$ Mpc and including winds around all 641
resolved galaxies in the box. The largest winds in this model extend
to $R_{wind} \sim 2 h^{-1}$ Mpc, requiring an average propagation
speed $V\sim 750 {\rm km}\;{\rm s}^{-1}(1{\rm Gyr}/t)$ for a wind duration $t$
and $h=0.65$.  Only the most extreme wind models come close to matching
the observational results.  Note, in particular that $1 h^{-1}$ Mpc
winds do not eliminate, or even drastically reduce, absorption at
$\Delta_r \le 1 h^{-1}$ Mpc because much of the absorption at these
separations in {\it redshift space} comes from infalling gas that is
further than $1 h^{-1}$ Mpc in {\it real space}.  These peculiar
velocity effects are also the reason that photoionization has such a
tiny impact; even for near maximal models, the ionization does not
strongly affect gas at such large distances.

\section{Statistical Fluke?}

Since the innermost data point comes from only three galaxy-los pairs,
we must also ask whether it could just be an anomalous statistical
fluctuation.  We have done a Monte Carlo calculation in which we draw
500 sets of 3 galaxies from our population of $z=3$ simulated galaxies
and compute the value of $\langle D\rangle$ at $\Delta_r\le0.5 h^{-1}$
Mpc for each 3-tuple. Figure 1d shows the cumulative distribution of
the flux decrement from these samples.  We see that our baseline
calculation can virtually never get to decrements as low as those
observed.  Even with a redshift error of $300 {\rm km}\;{\rm s}^{-1}$,
one sees decrements below 0.36 only $\sim5\%$ of the time.  For
the most extreme wind model, we find decrements as low as the
observed one, $\langle D \rangle =0.11$, $\sim5\%$ of the time. Despite the limited size of the
current data set, the observed ``LBG proximity effect'' stands as a
striking result, not easily explained.


\bibliographystyle{aipproc}

\begin{thebibliography}
\expandafter\ifx\csname natexlab\endcsname\relax\def\natexlab#1{#1}\fi
\providecommand{\enquote}[1]{``#1''}
\expandafter\ifx\csname url\endcsname\relax
  \def\url#1{\texttt{#1}}\fi
\expandafter\ifx\csname urlprefix\endcsname\relax\def\urlprefix{URL }\fi

\bibitem[Adelberger et al.(2002)]{adelberger02}
Adelberger, K.L., Steidel, C.C., Shapley, A.E., Pettini, M. 2002,
ApJ, in press, astro-ph/0210314


\bibitem[Kollmeier et al.(2002)]{kollmeier02}
Kollmeier, J.A., Weinberg, D.H., Dav\'{e}, R., Katz, N., 2002, ApJ, submitted,
astro-ph/0209563 


\bibitem[Croft et al.(2002)]{croft02}
Croft, R., Hernquist, L., Springel, V., Westover, M., White, M., 2002,
ApJ, \textbf{580}, 634

\bibitem[Bruscoli et al.(2002)]{bruscoli02}
Bruscoli, M., Ferrara, A., Marri, S., Schneider, R., Maselli, A.,
Rollinde, E., Aracil, B., astro-ph/0212126 

\bibitem[Steidel, Pettini, \& Adelberger(2001)]{steidel01}
Steidel, C.~C., Pettini, M., \& Adelberger, K.~L.\ 2001, ApJ, \textbf{546}, 665

\bibitem[Steidel et al.(2002)]{steidel02}
Steidel, C., Hunt, M., Shapley, A., Adelberger, K., Pettini, M.,
Dickinson, M., \& Giavalisco, M.\ 2002, ApJ, \textbf{576},
653


\bibitem[Pettini et al.(2002)]{pettini02}
Pettini, M., Rix, S.~A., Steidel, C.~C., Adelberger, K.~L., Hunt,
M.~P., \& Shapley, A.~E.\ 2002, ApJ, \textbf{569}, 742


\end{thebibliography}

\end{document}